**Fine-scale population structure analysis in *Armadillidium vulgare* (Isopoda: Oniscidea) reveals strong female philopatry**

Sylvine Durand, Frédéric Grandjean, Isabelle Giraud, Richard Cordaux, Sophie Beltran-Bech*, Nicolas Bech*

UMR CNRS 7267 ; Laboratoire Écologie and Biologie des Interactions ; équipe Écologie, Évolution, Symbiose ; Bâtiment B8-B35. 5 rue Albert Turpain TSA 51106, 86073 POITIERS CEDEX 9, France.

* co-last authors. These authors have contributed equally to this study.

Corresponding author: Sylvine Durand (durand.sylvine@gmail.com)

**Abstract**

In the last decades, dispersal studies have benefited from the use of molecular markers for detecting patterns differing between categories of individuals and have highlighted sex-biased dispersal in several species. To explain this phenomenon, several hypotheses implying mating systems, intrasexual competition or sex-related handicaps have been proposed. In this context, we investigated sex-biased dispersal in *Armadillidium vulgare*, a terrestrial isopod with a promiscuous mating system. As a proxy for effective dispersal, we performed a fine-scale investigation of the spatial genetic structure in males and females, using individuals originating from five sampling points located within 70 meters of each other. Based on microsatellite markers and spatial autocorrelation analyses, our results revealed that while males did not present a significant genetic structure at this geographic scale, females were significantly and genetically more similar to each other when they were collected in the same sampling point. As females invest more parental care than males in *A. vulgare*, but also because this species is promiscuous and males experience a high intrasexual competition, our results meet the predictions of most classical hypotheses for sex-biased dispersal. We suggest that widening dispersal studies to other isopods or crustaceans, differing in their ecology or mating system and displaying varying levels of parental care, might shed light on the processes underlying the evolution of sex-biased dispersal.

**Keywords**: philopatry, spatial genetic analysis, crustacean, parental care, mating system

**Introduction**

Dispersal is the process through which an organism contributes to gene flow by moving away from its natal population or current breeding site to another breeding site (Clobert *et al.*, 2001). This process is a major topic in biology due to its impact on species distribution, population structure and dynamics, as well as individual fitness through reproduction, growth and survival (Clobert *et al.*, 2001; Nathan, 2001). An abundant body of literature has highlighted an important variability in dispersal patterns between or within species according to their ecological requirements, environmental factors such as seasonality (Fies et *al.*, 2002), or individual characteristics such as age (Marvá & San Segundo, 2018) or sex (Trochet *et al*., 2016). Understanding the patterns of within-population spatial structure and their variation according to those factors promises to provide important insights into evolutionary and demographic strategies, which is crucial for population management in the context of population conservation or control (Jongejans *et al.*, 2008).

While dispersal patterns can be evaluated based on direct measurements of movements in the field (Nathan, 2001), for example using tracking devices (Kays *et al.*, 2015) or mark-recapture (Moore *et al.*, 2008), those methods have drawbacks that can be avoided by using genetic tools. Indeed, genetic approaches can be suited for cryptic or threatened species (Pérez-Portela *et al*., 2013 ; Gour *et al*., 2013, respectively) as well as for species whose small size makes them difficult to track using physical methods (Bilton *et al.,* 2001). Molecular markers can be used to estimate gene flow and migration rates, which are expected to be influenced by dispersal (Broquet & Petit, 2009; Legendre & Fortin, 2010). In particular, genetic methods have proven to be especially efficient in the context of sex-biased dispersal (Prugnolle & de Meeus, 2002; Kuhn *et al.*, 2017). Philopatry in one sex and dispersal in the alternate sex can indeed generate genetic discontinuities within populations (Chesser, 1991a; b), leading to different patterns of genetic structure between sexes.

Mechanisms responsible for the evolution of sex-biased dispersal are still highly debated and several hypotheses have been proposed based on both empirical evidence (reviewed in Trochet *et al.*, 2016) and theoretical work (Lehmann & Perrin, 2003, reviewed in Li & Kokko, 2018). Parameters related to mating systems were historically thought to be the main driver of sex-biased dispersal (Greenwood, 1980). For example, the sex experiencing the higher intrasexual competition is expected to disperse more (Dobson, 1982). Dispersal in one sex has also been suggested as an inbreeding avoidance strategy, leading to a decreased probability of encountering close relatives of the opposite sex (Pusey, 1987). The relative importance of such parameters remains highly debated today (Dobson, 2013). Other authors have underlined the importance of unequal dispersal cost between sexes, the sex experiencing lower costs being expected to disperse more (Gros *et al*., 2008, Bonte *et al*., 2012). In this context, the presence of handicaps in one sex, such as the presence of expensive parental care, has recently been presented as a valuable alternative hypothesis by Trochet *et al.* (2016). Under the latter scenario, dispersal should be more costly for the sex that invests the most in parental care and thus counter-selected.

The parental care-handicap hypothesis may be particularly relevant to explain sex-biased dispersal patterns in crustaceans. In these animals, females carry their offspring for variable amounts of time (Sastry, 1983), and increased predation risk associated to locomotion costs have been suggested for females in several species such as copepods (Svetlichny *et al.,* 2017), amphipods (Williams *et al.,* 2016) and isopods (Suzuki and Futami, 2018). Such female-related costs could lead to male-biased dispersal. Although information about sex-specific dispersal is scarce concerning crustaceans, this prediction has been verified in the crayfish *Pacifastacus leniusculus* (Hudina *et al.*, 2012; Wutz & Geist, 2013). Surprisingly, female-biased dispersal has been recorded in some amphipods (*Paracorophium* spp. (Stevens *et al.*, 2006) and *Corophium volutator* (Bringloe *et al.*, 2013) and a shrimp (*Aristeus antennatus* (Cannas *et al.*, 2012)). Other factors such as mating systems may then contribute to shape sex-biased dispersal in these organisms. These contradictory results illustrate the lack of knowledge concerning crustaceans, and more generally invertebrates, and highlight the need to perform integrative studies accounting for the various life-history traits potentially implied in dispersal.

Here we propose to test for sex-biased dispersal in the terrestrial isopod *Armadillidium vulgare*. In this crustacean, females display important maternal care by incubating their offspring during one month in a ventral pouch to allow their development (Bech *et al.*, 2017), leading to high costs for females (Kight & Ozga, 2001; Appel *et al.*, 2011). Females usually produce two to three broods per year between April and October (Vandel, 1962). By contrast, males only invest in sperm production. Females in several terrestrial isopod species have been shown to reproduce with several males (Johnson, 1982; Sassaman, 1978; Moreau *et al.*, 2002) and males can mate with several females (Moreau and Rigaud, 2003). In particular, *A. vulgare* presents a promiscuous system, as females can produce broods with up to 7 fathers in the wild (Valette *et al.*, 2017). Due to small size and regular moulting, genetic tools appear particularly relevant to study dispersal in this species. Such tools have already been used to study population structure in terrestrial isopods (Verne *et al.*, 2012), and highlighted notably a significant isolation by distance (IBD) in *A. vulgare*. However, this study was performed over a range of several tens of kilometers, which is likely way over the dispersal capacities of these small animals. A study at a finer scale, more compatible with the species characteristics, is then required to study spatial structure with precision. Such an approach would also allow for an estimation of fine-scale differences in dispersal between sexes. Indeed, according to Goudet *et al.* (2002), a genetic signature of a sex-biased dispersal cannot be detected in highly structured populations, often occurring at a large geographical scale. Thus, we carried out our study at individual scale and over a range of a few tens of meters to assess for potential sex-biased dispersal. As the magnitude and direction of sex bias in dispersal are difficult to highlight directly following movements in the field, we employed an indirect molecular approach (*i.e.* microsatellite markers) to infer sex-specific genetic structure from the spatial distribution of alleles in the gregarious species *A. vulgare* (Goudet & Waser, 2002).

**Materials and methods**

**Sampling and molecular analyses**

We collected a population of 53 *A. vulgare* individuals in La Crèche (France). This species is gregarious and often found in moist and dark habitats such as under trees or rocks. For this reason, individuals were collected on 5 different sampling points displaying landscape features allowing woodlouse aggregation and located in grassland, a suitable continuous habitat for *A. vulgare* movement. For each sampling point, we recorded geographic coordinates using a Global Positioning System (centroid of sampling points: 46° 21' 38" N, 0° 18' 20" W). These sampling points were contained within a 5200 m² area (Figure 1). Among collected individuals, there were 35 females and 18 males. Adults of approximately the same size (*i.e.* same age, approximately one year), were collected in April 2017, at the beginning of the reproductive season (Vandel, 1962). Even though the exact timing of dispersal is currently unknown in this species, sampling right before reproduction allowed us to focus on "natal dispersal" as defined by Greenwood, *i.e.* "movement from birth site to first breeding site" (Greenwood, 1980). Hence, here we concentrated on the minimal dispersal that contributes to gene flow, even if some dispersal may occur again before subsequent reproductive events.

From each individual, we extracted total genomic DNA from the head and all 14 legs using the DNeasy Blood and Tissue kit (QIAGEN, Basel, Switzerland) according to the manufacturer protocol. The endosymbiont *Wolbachia* can be present in some populations and is known to affect several aspects of *A. vulgare* biology, especially regarding reproduction or behaviour. As it may eventually affect dispersal abilities, we verified its absence in all individuals by means of PCR, using the molecular marker *wsp* (Cordaux *et al.*, 2012), thus ruling out any potential confounding effect due to *Wolbachia* in our study. We genotyped these individuals using 9 microsatellite markers previously described in this species (Verne *et al.*, 2006; Giraud *et al.*, 2013) (Table 1) using a Multiplex PCR kit (QIAGEN, Basel, Switzerland). All PCRs were carried out according to the manufacturer's standard microsatellite amplification protocol in a final volume of 10 μL and an annealing temperature of 57 °C, as described in Durand *et al.* (2015). PCR product separation was then performed by electrophoresis on an automated sequencer (ABI) by Genoscreen (Lille, France). Fragment size was determined using GeneMapper version 3.7 (Applied Biosystems).

**Figure 1:** Map of study area and associated sampling points: point 1 (F=5; M=6), point 2 (F=5;M=8), point 3 (F=12; M=1), point 4 (F=3; M=2) and point 5 (F=10; M=1) (F=sample size of females; M=sample size of males).

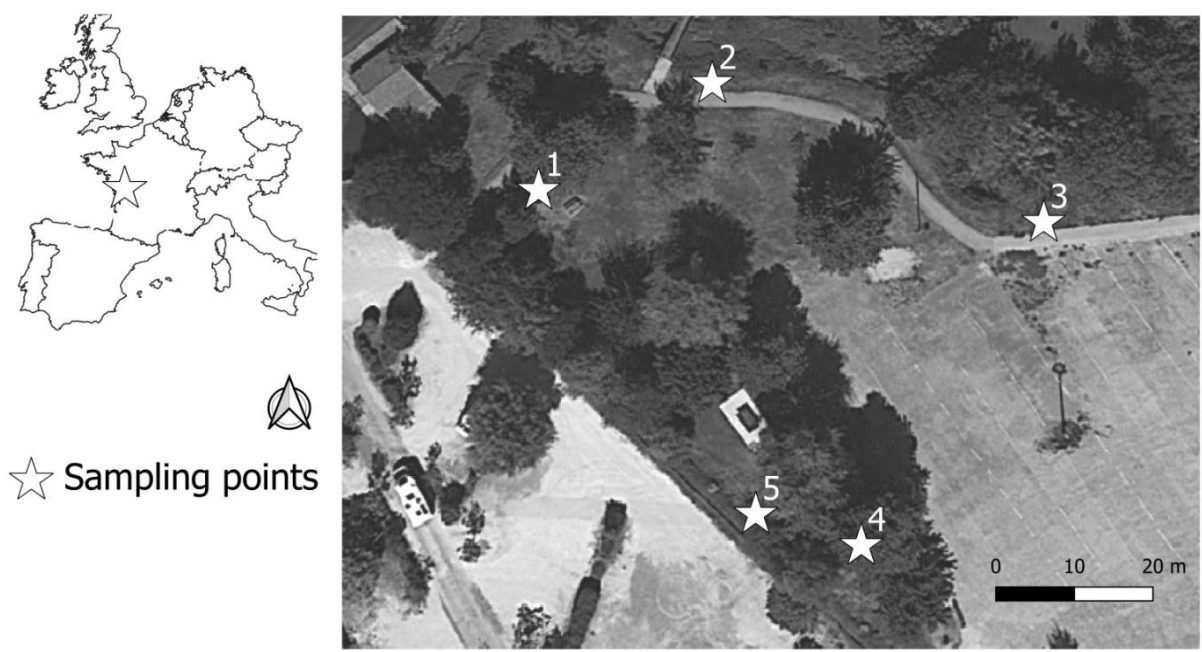

**Global genetic analyses**

For global genetic analyses, we considered only one population including all 53 individuals. This was supported by the lack of significant genetic structure observed between the different sampling points (Supplementary file 1). For the unique population including all individuals collected from the five different sampling points, we used MICROCHECKER version 2.2.3 (Van Oosterhout *et al.*, 2004) to detect signs of null alleles or scoring errors due to stuttering. We tested linkage disequilibrium and departures from Hardy-Weinberg expectations using FSTAT 2.9.3.2 (10 000 permutations) (Goudet, 2001). We adjusted the level of significance for multiple tests using the Bonferroni correction. We estimated polymorphism for each locus using allelic richness (AR), observed and expected heterozygosity (Ho, He respectively) and the $F_{IS}$ for all individuals, using FSTAT v.2.9.3.2 (Goudet, 2001).

**Spatial and genetic autocorrelations**

At the individual scale (including all males and females), we tested the relationship between Euclidean geographic distances (computed from the geographic coordinates of our sampling points) and genetic distances with a Mantel test (with 9999 permutations), using GENALEX software v 6.2 (Peakall & Smouse, 2006, 2012). Although this test is often used to test relationships between genetic data and the spatial pattern of sampling locations, it is known to underestimate the variation explained by the spatial structure (Legendre & Fortin, 2010). Thus, to complement the Mantel test, we used an alternative and more powerful spatial autocorrelation method implemented in the GENALEX software v.6.2 combining spatial data and multilocus genotypes. This analysis generated an autocorrelation coefficient *r* using the matrices of pairwise geographic distances and of pairwise genetic distances for all individual pairs. We computed pairwise genetic distances using the Codom-genotypic option provided by the GENALEX software. These genetic distances are based on the number of alleles shared by both individuals and their respective heterozygosity level (Peakall & Smouse, 2006). The autocorrelation coefficient *r* was calculated for different geographical distance classes (here 7 classes separated by a 10 m interval, from 0 to 70 m), and ranges from -1 to +1 (Peakall *et al.*, 2003). Thereby, a positive autocorrelation coefficient only for the first lowest distance classes would reflect high local genetic similarity and thus restricted dispersal capacities. We performed this analysis separately for males (153 pairwise comparisons) and females (595 pairwise comparisons) to test for sex-biased dispersal, using the 'single pop' option implemented in the GENALEX software v.6.2. The error around the *r* coefficient for each distance class was estimated by 9999 bootstraps. The *r* values were then compared to the null hypothesis of no autocorrelation (*i.e. r* = 0), for which a 95% confidence interval was determined by 9999 random permutations (Peakall *et al.*, 2003; Neville *et al.*, 2006). Results were plotted in correlograms displaying variations of the *r* coefficient according to different geographical distance classes. As suggested in Peakall *et al.* (2003), we considered a significant autocorrelation for a given distance class when both the estimated *r* coefficient was outside the 95% confidence interval and the *r* error bar did not cross the x axis (*r* = 0). We also tested specifically for the presence of significant positive autocorrelation using a one-tailed test as in Peakall *et al.* (2003), because positive autocorrelation is predicted at short distances under restricted dispersal. Moreover, one may argue that the small sample size in our male analysis (only 18) may prevent the detection of a significant spatial genetic structure. Thus, we ran other autocorrelations analyses including only 18 females randomly sampled amongst 35 to determine whether results were consistent between sample sizes in females (18 or 35). This analysis on 18 randomly sampled females was performed 10 times independently to examine the extent to which an analysis run on only 18 individuals could be trusted (Supplementary file 2).

**Results**

**Global genetic analyses of the population**

Among the 9 microsatellites markers, we detected no evidence for null allele, linkage disequilibrium (adjusted significance threshold P=0.0014 and all P>0.019) and departure from Hardy–Weinberg equilibrium (adjusted significance threshold P=0.0056 and all P>0.172) (Table 1). All microsatellite loci were polymorphic with an allelic richness ranging from 3 to 20, an observed heterozygosity ranging from 0.373 to 0.894 and an expected heterozygosity ranging from 0.409 to 0.925 (Table 1).

**Spatial genetic structure and sex-biased dispersal**

Results from the Mantel test including all individuals revealed a significant correlation between genetic and geographical distances, suggesting a population structured in an IBD pattern ($r^2$=0.021; P<0.001). At a finer scale, the spatial autocorrelation analysis performed on females revealed a significant and positive autocorrelation in the first distance class (*i.e.* <10 m), but not in higher distance classes. This result indicated that females were more genetically similar to females from the same sampling point than to females collected more than 10 m away (Figure 2A, supplementary file 1). Conversely, even in the shortest distance class, the male autocorrelogram did not reveal any significant genetic autocorrelation. Indeed, as all *r* values were contained within the 95% confidence interval (Figure 2B), correlograms for males suggested that pairwise genetic distances were completely independent from pairwise geographic distances.

**Figure 2:** Correlograms displaying variation of the *r* coefficient according to geographical distance classes for *Armadillidium vulgare* (A) females and (B) males. The number of pairwise comparisons used in the computation of *r* for each distance class is indicated. Grey areas correspond to the 95% confidence interval for the null hypothesis of no spatial structure (no autocorrelation, *r*=0). * indicates significant positive autocorrelation.

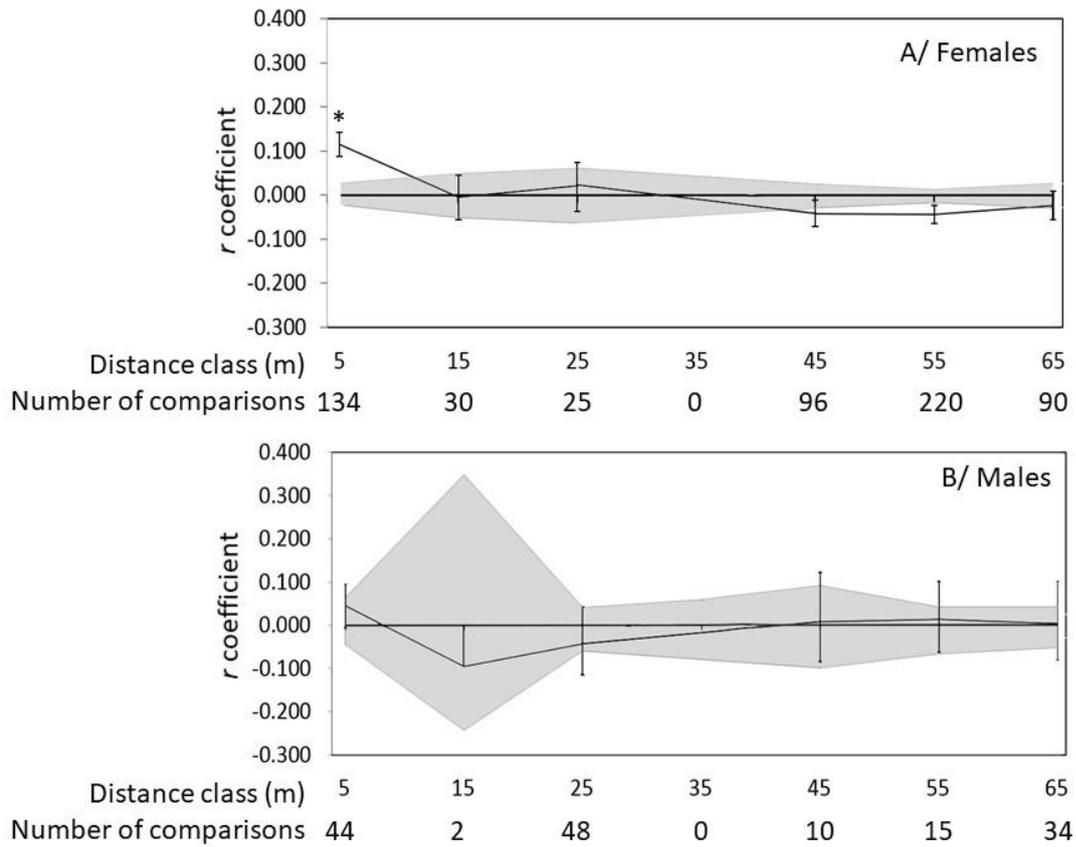

**Discussion**

We investigated the spatial genetic structure, used as a proxy for effective dispersal, within an *Armadillidium vulgare* population. According to the presumed restricted dispersal abilities of this small terrestrial invertebrate, we performed our analyses at a fine scale along a gradient of geographical distances (maximum 70 m). IBD was detected, which suggests a limited gene flow within the small study area. These results corroborate and extend previous findings obtained on the same species over a range of several tens of kilometers (Verne *et al.*, 2012).

Interestingly, a sex-related genetic structure was detected using autocorrelation analyses. Females were more genetically similar to other females within the same sampling point, but not in higher distances classes, suggesting a spatial genetic structure for females at a very short scale. No significant genetic autocorrelation was detected for males along the gradient of distances, suggesting a constant genetic similarity between males regardless of the distance between them. These results are robust to sample size because similar results were obtained for females using all 35 individuals or subsamplings of 18 individuals (*i.e.* corresponding to the male sample size) (supplementary file 2). Thus, with this design and this microsatellite set, a sample size of 18 individuals is adequate to detect a signal such as the one detected in females. This indicates either that males present no significant spatial structure at all, or that they present a weaker signal than in females, requiring more than 18 individuals to be detected. Whatever the case, we can claim that females are more philopatric than males.

Our results on both sexes are consistent with previous observations on *A. vulgare*. First, females have been suggested to be more grouped than wandering males at the beginning of the reproductive season (Caubet *et al.*, 1998), this differential mobility being in line with female philopatry. This may also explain the female-biased sex-ratio observed in our sampling: if males disperse more than females, they may be less present in sampled aggregates. Regarding genetic observations, the strong mitochondrial structure and the weak nuclear genetic structure observed at a scale of several kilometers (Verne *et al.*, 2012) are respectively consistent with the local female structure and the male-related local gene flow observed here. This male-mediated gene flow might also allow connecting sampling points by mixing genes at each generation for both sexes, explaining the weak overall genetic structure we found (supplementary file 1). As a perspective, genetic structure could also be studied at a larger scale beyond male dispersal capacities (*e.g.* potentially several hundred meters) to test for the presence of significant genetic structure in males and to estimate its extent. Enlarging the study area might also allow for inclusion of a more diverse landscape, which could enable evaluation of the impact of potential barriers on gene flow in this terrestrial invertebrate. However, we think that our result can be generalized to other populations at the same local scale in other suitable habitats.

Globally, our results suggest strong female philopatry and male dispersal at a small scale in *A. vulgare*. Several non-mutually exclusive hypotheses can explain these results. The local mate competition hypothesis (Dobson, 1982) postulates that the sex suffering the most from intrasexual competition should disperse more. As males possess a higher mating capacity (Moreau & Rigaud, 2003) than females (Moreau *et al.,* 2002), they are likely to undergo higher intrasexual competition in terrestrial isopods which generally display a balanced operational sex-ratio (Moreau & Rigaud, 2000), potentially leading to male-biased dispersal. Inbreeding avoidance has also been proposed as a determinant for sex-biased dispersal (Pusey & Wolf, 1996). In particular, mathematical predictions suggest that female choice for inbreeding avoidance might promote male dispersal if inbreeding costs are high (Lehmann & Perrin, 2003). In *A. vulgare*, inbreeding avoidance through mate choice has been suggested in both sexes (Durand *et al.,* 2015; 2017), and even though inbreeding costs remain to be fully evaluated, a decrease in offspring number for similar parents has been highlighted (Durand *et al.,* 2017), potentially through mortality events around birth (Durand *et al*., 2018). The sex-biased dispersal observed in this study could thus be an additional mechanism allowing avoiding costly inbreeding in this species. On the other hand, because females incubate their offspring in a ventral pouch for one month (Bech *et al.,* 2017) and bear significant costs upon locomotion (Kight & Ozga, 2001, Suzuki & Futami, 2018) and likely food intake (Appel *et al.,* 2011), our results are also in line with the predictions of the handicap hypothesis proposed by Trochet *et al*. (2016).

We studied sex-biased dispersal at a fine spatial scale using molecular tools for the first time on a terrestrial isopod species. Our results support female philopatry in *A. vulgare*, whereas no structure was detected in males. Information on dispersal for both sexes is available for only one other isopod species, the desert isopod *Hemilepistus reaumuri*. This species presents a dispersal pattern different than that observed in *A. vulgare*, as mark-recapture methods showed no sex-difference in travelled distances (Baker, 2004). Interestingly, this species displays both a monogamous mating system and biparental care (Linsenmair, 1984), as opposed to *A. vulgare*, which is characterized by a promiscuous mating system (Moreau & Rigaud, 2003; Valette *et al.,* 2017) and exclusive maternal care. It is then not surprising to observe different dispersal patterns across terrestrial isopods given the diversity in their ecology.

The diversity of dispersal patterns illustrates the need to perform comparative studies between phylogenetically related species presenting a high variability in ecology, morphology and mating systems, to evaluate the relative importance of each suggested determinant for sex-biased dispersal. While this has thoroughly been performed on vertebrates, especially birds and mammals (Greenwood, 1980; Dobson, 1982; Dobson, 2013), invertebrates remain quite left behind (Downey *et al.,* 2015). However, crustaceans, and especially isopods, appear to be particularly relevant to perform such studies because of their diversity in ecology (terrestrial *vs* aquatic lifestyles), social structure (different degrees of gregariousness; Broly *et al.,* 2013) and mating systems (Lefebvre, 2002).

**Authors' contributions**

FG, RC, SBB and NB conceived and designed the study. FG and NB sampled animals on the field. SD and IG performed molecular work. SD, SBB and NB performed the analyses and drafted the manuscript. All authors contributed to the manuscript and approved its content. The preprint of this work has been deposited on the repository arXiv ([arXiv:1807.03059](arXiv:1807.03059)).

**Acknowledgements**

We thank two anonymous reviewers for constructive comments on the manuscript. This work was funded by Agence Nationale de la Recherche grant ANR-15-CE32-0006-01 (CytoSexDet) to R.C., the PEPS-CNRS 2019 (CORECO) to N.B., the 2015-2020 State-Region Planning Contracts (CPER), the European Regional Development Fund (FEDER), the partnership arrangements in ecology and the environment (DIPEE) and intramural funds from the Centre National de la Recherche Scientifique, and the University of Poitiers.

|          |          | All individuals | | | |
| --- | --- | --- | --- | --- | --- |
| Loci | Multiplex | Ar | Ho | He | $F_{IS}$ |
| **Av2†** | 1 | 5.000 | 0.569 | 0.692 | 0.046 |
| **Av4†** | 1 | 4.920 | 0.620 | 0.621 | 0.066 |
| **Av5†** | 1 | 13.594 | 0.894 | 0.796 | -0.030 |
| **Av3†** | 2 | 3.936 | 0.823 | 0.409 | 0.052 |
| **Av6†** | 2 | 13.745 | 0.540 | 0.646 | 0.020 |
| **AV0018*** | 3 | 20.000 | 0.667 | 0.925 | 0.037 |
| **AV0032*** | 3 | 4.997 | 0.373 | 0.507 | 0.114 |
| **AV0056*** | 3 | 5.000 | 0.460 | 0.575 | 0.077 |
| **AV0063*** | 3 | 3.000 | 0.480 | 0.429 | -0.095 |
| **All loci** |  | 8.244 | 0.603 | 0.622 | 0.032 |

**Table 1**: Characterization of the 9 microsatellite loci amplified in the *Armadillidium vulgare* population from La Crèche, France. References of microsatellite molecular markers [†: Verne et al. (2006); *: Giraud et al. (2013)], multiplex number, allelic richness (Ar), observed heterozygosity (Ho), expected heterozygosity (He) and $F_{IS}$ are shown. No $F_{IS}$ value was significantly different from 0.

# Supplementary file 1: Do all sampling points belong to a single population?

Considering the geographical proximity between our sampling points, we tested herein if all individuals could be gathered together into a single genetic population.

**METHODS**: We estimated the pairwise $F_{ST}$ values according to Weir and Cockerham (1984) between the five sampling points. We computed these values using the software FSTAT 2.9.3.2 (Goudet, 2001) and tested their significance using the sequential Bonferroni correction to adjust the level of significance for multiple testing (Rice, 1989).

Moreover, we also used an individual-based approach to estimate the number of panmictic groups. Specifically, we used the Bayesian approach implemented in STRUCTURE version 2.2 software (Pritchard *et al.*, 2000) which allows estimating both the number of genetic groups (*i.e.* K clusters) and the admixture coefficient of individuals to be assigned to the estimated clusters (Pritchard *et al.*, 2000). We chose the correlated allele frequencies among populations and admixture model. Each simulation (with K from 1 to 10) was replicated 20 times as recommended by (Evanno *et al.*, 2005), with a $10^4$ burn-in period followed by $10^6$ steps. To determine the number of independent genetic populations (K), we compared the mean likelihood and variance between our different K values computed from the 15 independent runs using STRUCTURE HARVESTER version 0.6.1 (Earl & vonHoldt, 2012).

**Results**:

We did not detect any evidence for genetic structure with the F-statistics, which revealed no significant genetic differentiation between individuals belonging to the sampling points (mean $F_{ST}$ = 0.010, all p-values > 0.005) (Table S1).

**Table S1**: $F_{ST}$ values for each pairwise comparison between sampling points (below diagonal), and associated p-values (above diagonal). There was no significant genetic differentiation (indicative adjusted nominal level (5%) for multiple comparisons: 0.005).

|  | Sampling point 1 | Sampling point 2 | Sampling point 3 | Sampling point 4 | Sampling point 5 |
|---|---|---|---|---|---|
| **Sampling point 1** | - | 0.295 | 0.17 | 0.16 | 0.185 |
| **Sampling point 2** | 0.0241 | - | 0.075 | 0.42 | 0.56 |
| **Sampling point 3** | 0.0035 | 0.0257 | - | 0.085 | 0.475 |
| **Sampling point 4** | 0.0282 | 0.0089 | 0.0189 | - | 0.25 |
| **Sampling point 5** | 0.0009 | -0.0063 | -0.0011 | -0.0009 | - |

In agreement with this result, the weak genetic structure was also supported by the Bayesian clustering method indicating a lack of genetic structure with the highest mean likelihood for only one genetic cluster (K=1) (figure S1). Combined with similar admixture coefficients inferred for each individual (results not shown), this suggests a very strong genetic homogeneity within the whole sampling and so allows us considering that all individuals can be gathered together into a single genetic population.

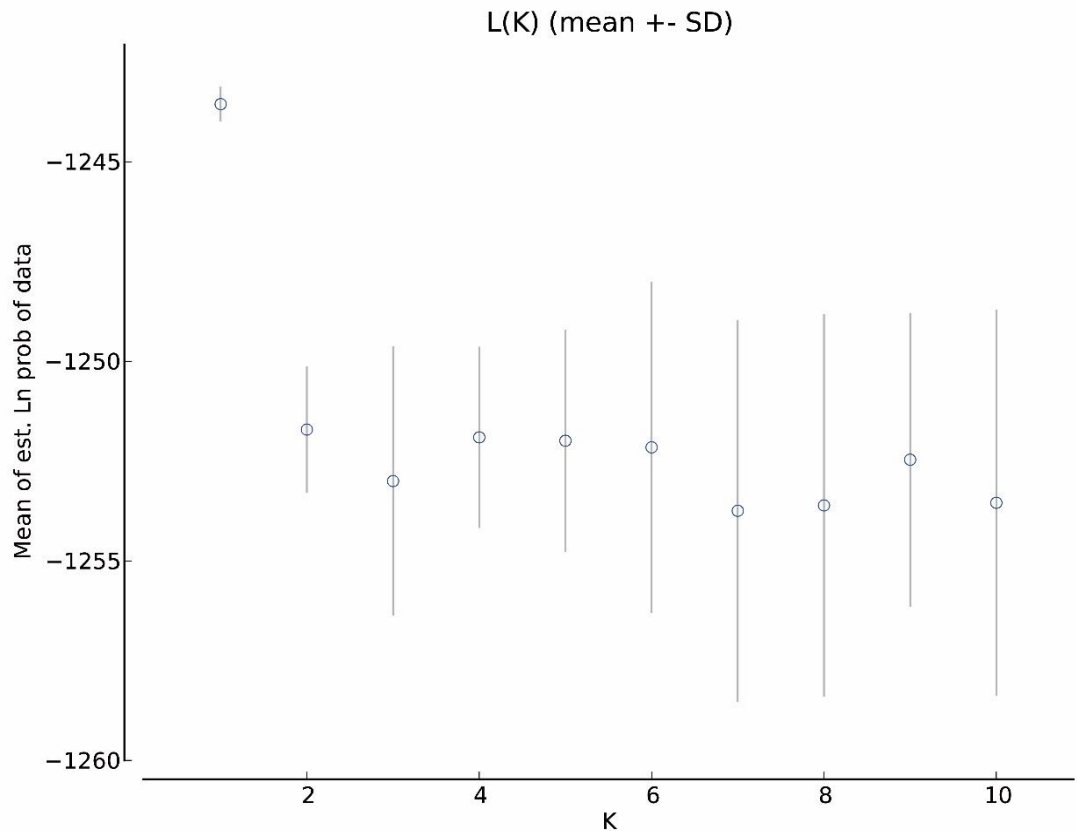

**Figure S1:** Plot of STRUCTURE results showing mean likelihood (along with their variance across the 20 replicates) per number of simulated genetic clusters (K). The highest mean likelihood is obtained for K=1 suggesting a lack of genetic structure.

# Supplementary file 2: Results from 10 spatial autocorrelation analyses performed on 18 randomly selected females.

At a fine scale, the spatial autocorrelation analysis performed on females revealed a significant and positive autocorrelation in the first distance class (*i.e.* <10 m). This result indicated that females were more genetically similar to females from the same sampling point than to females collected more than 10 m further (Table S2 below). Conversely, even in the shortest distance class, the male analysis did not reveal any significant genetic autocorrelation. Thus, results suggested that pairwise genetic distances between males were completely independent from pairwise geographic distances.

|  |  | Distance Class (Mid Point) | | | | | | |
|---|---|---|---|---|---|---|---|---|
|  |  | 5 | 15 | 25 | 35 | 45 | 55 | 65 |
| **Results from all 35 females** | n | 134 | 30 | 25 | 0 | 96 | 220 | 90 |
|  | r | 0.110 | 0.002 | 0.021 | NA | -0.044 | -0.039 | -0.027 |
|  | U | 0.026 | 0.048 | 0.06 | NA | 0.024 | 0.013 | 0.025 |
|  | L | -0.022 | -0.053 | -0.061 | NA | -0.028 | -0.017 | -0.028 |
|  | p | ***0.000*** | 0.478 | 0.245 | NA | 0.998 | 1.000 | 0.972 |
| **Results from all 18 males** | n | 44 | 2 | 48 | 0 | 10 | 15 | 34 |
|  | r | 0.044 | -0.096 | -0.044 | NA | 0.008 | 0.013 | 0.003 |
|  | U | 0.065 | 0.347 | 0.040 | NA | 0.091 | 0.042 | 0.042 |
|  | L | -0.044 | -0.242 | -0.058 | NA | -0.098 | -0.065 | -0.052 |
|  | p | 0.068 | 0.751 | 0.948 | NA | 0.470 | 0.282 | 0.481 |

**Table S2: Results from spatial autocorrelation analyses performed separately on all females and males.** These analyses were carried out using the 'single pop' option implemented in the GENALEX software v.6.2. We computed the autocorrelation coefficient *r* for each geographical distance class (*i.e.* 7 classes separated by a 10 m interval, from 0 to 70 m). This coefficient ranged from -1 to +1 (Peakall *et al.*, 2003). A 95% confidence interval was determined for the null hypothesis of no autocorrelation (*r*=0) by 9999 random permutations (Peakall *et al.*, 2003; Neville *et al.*, 2006). We also tested specifically for positive autocorrelation using a unilateral test. n : number of pairwise comparisons; *r* = autocorrelation coefficient ; U and L : respectively upper and lower bounds of the 95% confidence interval for the null hypothesis of no spatial structure (*i.e. r* =0); p : p-value after unilateral test for positive autocorrelation (bold and italics indicate significant p-values). A: not available in this distance class.

As the number of males was only 18, it could be argued that the absence of spatial genetic structure results from the low sample size. To test the robustness of our results, we randomly selected 18 females among the 35 to implement them in a new run of spatial autocorrelation analysis. This procedure was replicated 10 times and allowed us to determine whether the new results were consistent with the one previously obtained with 35 females.

The 10 simulations, based on 18 randomly selected females, yielded results (and related statistics) similar to those obtained on 35 females. Thus, given our dataset, the spatial autocorrelation analysis was able to detect significant results even with 18 individuals (table S3 below). So, we assumed that our results concerning males were robust.

|  |  | Distance Class (Mid Point) | | | | | | |
|---|---|---|---|---|---|---|---|---|
|  |  | 5 | 15 | 25 | 35 | 45 | 55 | 65 |
| **Simulation 1** | n | 31 | 6 | 9 | 0 | 20 | 66 | 21 |
|  | *r* | 0.132 | -0.090 | 0.039 | NA | 0.014 | -0.046 | -0.047 |
|  | U | 0.056 | 0.120 | 0.102 | NA | 0.059 | 0.024 | 0.057 |
|  | L | -0.047 | -0.106 | -0.096 | NA | -0.063 | -0.031 | -0.060 |
|  | p | ***0.000*** | 0.950 | 0.216 | NA | 0.335 | 0.998 | 0.940 |
| **Simulation 2** | n | 30 | 12 | 6 | 0 | 20 | 60 | 25 |
|  | *r* | 0.104 | 0.018 | 0.038 | NA | -0.015 | -0.040 | -0.035 |
|  | U | 0.055 | 0.076 | 0.121 | NA | 0.055 | 0.025 | 0.048 |
|  | L | -0.046 | -0.074 | -0.116 | NA | -0.058 | -0.032 | -0.052 |
|  | **p** | ***0.001*** | 0.323 | 0.260 | NA | 0.713 | 0.990 | 0.913 |
|  | n | 35 | 14 | 3 | 0 | 25 | 63 | 13 |
|  | r | 0.120 | -0.027 | 0.003 | NA | -0.031 | -0.059 | 0.060 |
| **Simulation 3** | U | 0.050 | 0.067 | 0.200 | NA | 0.050 | 0.025 | 0.078 |
|  | L | -0.041 | -0.068 | -0.164 | NA | -0.054 | -0.032 | -0.072 |
|  | **p** | ***0.000*** | 0.794 | 0.463 | NA | 0.886 | 0.999 | 0.060 |
|  | n | 37 | 9 | 3 | 0 | 32 | 36 | 36 |
|  | r | 0.114 | 0.036 | 0.010 | NA | -0.072 | -0.056 | 0.002 |
| **Simulation 4** | U | 0.044 | 0.097 | 0.185 | NA | 0.038 | 0.031 | 0.035 |
|  | L | -0.039 | -0.096 | -0.169 | NA | -0.044 | -0.035 | -0.038 |
|  | **p** | ***0.000*** | 0.225 | 0.450 | NA | 0.999 | 0.999 | 0.471 |
|  | n | 30 | 6 | 16 | 0 | 15 | 66 | 20 |
|  | r | 0.110 | 0.089 | 0.020 | NA | -0.033 | -0.053 | -0.005 |
| **Simulation 5** | U | 0.060 | 0.120 | 0.070 | NA | 0.071 | 0.024 | 0.060 |
|  | L | -0.052 | -0.115 | -0.076 | NA | -0.074 | -0.032 | -0.064 |
|  | **p** | ***0.001*** | 0.072 | 0.294 | NA | 0.817 | 0.998 | 0.583 |
|  | n | 38 | 12 | 2 | 0 | 28 | 60 | 13 |
|  | r | 0.097 | -0.018 | 0.123 | NA | -0.023 | -0.047 | -0.013 |
| **Simulation 6** | U | 0.045 | 0.076 | 0.227 | NA | 0.044 | 0.025 | 0.071 |
|  | L | -0.038 | -0.074 | -0.214 | NA | -0.048 | -0.030 | -0.067 |
|  | **p** | ***0.000*** | 0.699 | 0.136 | NA | 0.845 | 0.998 | 0.659 |
|  | n | 34 | 6 | 6 | 0 | 18 | 66 | 23 |
|  | r | 0.125 | 0.027 | -0.001 | NA | -0.049 | -0.054 | 0.004 |
| **Simulation 7** | U | 0.056 | 0.120 | 0.130 | NA | 0.061 | 0.025 | 0.053 |
|  | L | -0.046 | -0.114 | -0.128 | NA | -0.067 | -0.034 | -0.061 |
|  | **p** | ***0.000*** | 0.324 | 0.513 | NA | 0.932 | 0.998 | 0.455 |
|  | n | 31 | 6 | 9 | 0 | 20 | 66 | 21 |
|  | r | 0.102 | 0.060 | 0.030 | NA | -0.079 | -0.015 | -0.050 |
| **Simulation 8** | U | 0.054 | 0.114 | 0.095 | NA | 0.056 | 0.023 | 0.053 |
|  | L | -0.046 | -0.101 | -0.090 | NA | -0.060 | -0.029 | -0.057 |
|  | **p** | ***0.000*** | 0.146 | 0.275 | NA | 0.995 | 0.874 | 0.958 |
|  | n | 29 | 15 | 4 | 0 | 35 | 50 | 20 |
|  | r | 0.089 | 0.055 | 0.018 | NA | -0.026 | -0.035 | -0.043 |
| **Simulation 9** | U | 0.049 | 0.062 | 0.137 | NA | 0.036 | 0.027 | 0.051 |
|  | L | -0.046 | -0.061 | -0.123 | NA | -0.038 | -0.029 | -0.052 |
|  | **p** | ***0.000*** | 0.039 | 0.387 | NA | 0.913 | 0.991 | 0.946 |
|  | n | 38 | 3 | 8 | 0 | 40 | 42 | 22 |
|  | r | 0.093 | 0.221 | 0.039 | NA | -0.078 | -0.025 | -0.012 |
| **Simulation 10** | U | 0.045 | 0.182 | 0.100 | NA | 0.033 | 0.024 | 0.050 |
|  | L | -0.035 | -0.157 | -0.100 | NA | -0.041 | -0.028 | -0.051 |
|  | **p** | ***0.000*** | 0.010 | 0.216 | NA | 1.000 | 0.958 | 0.693 |

**Table S3: Results from the 10 spatial autocorrelation analyses performed on 18 randomly selected females.** These analyses were carried out using the 'single pop' option implemented in the GENALEX software v.6.2. We computed the autocorrelation coefficient *r* for different geographical distance classes (*i.e.* 7 classes separated by a 10 m interval, from 0 to 70 m). This coefficient ranged from -1 to +1 (Peakall *et al.*, 2003). A 95% confidence interval was determined for the null hypothesis of no autocorrelation ($r$=0) by 9999 random permutations (Peakall *et al.*, 2003; Neville *et al.*, 2006). We also tested specifically for positive autocorrelation using a unilateral test. n : number of pairwise comparisons; $r$ = autocorrelation coefficient ; U and L : respectively upper and lower bounds of the 95% confidence interval for the null hypothesis of no spatial structure (*i.e.* $r$ =0); p : p-value after unilateral test for positive autocorrelation (bold and italics indicate significant p-values). NA: not available in this distance class.